\documentclass[
    ,final            % use final for the camera ready runs
  ]
  {aipproc}

\layoutstyle{6x9}
\def\roughly#1{\raise.3ex\hbox{$#1$\kern-.75em\lower1ex\hbox{$\sim$}}}
\newcommand{\AP}{{\alpha^{\prime}}}
\newcommand{\pd}{\partial}
\usepackage{amsmath}
\usepackage{amssymb}
\usepackage{graphicx}
\begin{document}

\title{Stringy Model of Cosmological Dark Energy}


\author{Irina Ya. Aref'eva\footnote{Email: arefeva@mi.ras.ru}}{
  address={Steklov Mathematical Institute of Russian
Academy of Sciences,\\ Gubkin st., 8, 119991, Moscow, Russia} }

\begin{abstract}
A string field theory (SFT) nonlocal model of the cosmological dark energy  providing $w
< -1$ is briefly surveyed. We summarize recent developments and open problems, as well as
point out some theoretical issues related with  others applications of the SFT nonlocal
models in cosmology, in particular, in inflation  and cosmological singularity.
 \end{abstract}

\maketitle

%%%%%%%%%%%%%%%%%%%%%%%%%%%%%%%%%%%%%%%%%%%%
%% MAINMATTER
%%%%%%%%%%%%%%%%%%%%%%%%%%%%%%%%%%%%%%%%%%%%

\section{Introduction}

The origin of the dark energy (DE) is still a fascinating  puzzle.
Present cosmological observations do not exclude an evolving  DE
state parameter $w$. According recent data (U.Seljak, Pascos, July 07)
$$w=-1.04\pm 0.06.$$
 There are
two questions to experimental data:
\begin{itemize}
\item
Can we rule out a dynamical DE?
\item Can we rule out $w<-1$?
\end{itemize}
 Recent data are not
enough to answer these two questions and moreover, within the next few years
answers on these two questions will not  accessible
 as it has  previously expected.
However one might wonder whether there is a room for $w<-1$ in a theory.

Dark energy models with the state parameter $w < -1$ violate the
null energy condition (NEC). All local models realizing the NEC
violation  are unstable and violate usual physical requirements.
 To provide $w <
-1$
 a string field theory (SFT) nonlocal DE model has been proposed \cite{IA}.

 In this SFT nonlocal model our Universe is considered as a D3
non-BPS brane embedded in the 10 dimensional space-time. The role of
the dark energy plays the Neveu-Schwarz (NS) string tachyon leaving in GSO$-$
sector. The tachyon action is dictated by  the cubic fermionic SFT \cite{AMZ,NPB}
 and it is  nonlocal due to string   effects \cite{review-SFT}.

 We  postulate  a minimal form of  the tachyon interaction with gravity. In the spatially flat FRW metric
the model is described by a system of two nonlinear nonlocal equations for the tachyon field
 and the Hubble parameter.
 The corresponding potential has perturbative and nonperturbative minima.
A transition from a perturbative vacuum to a non-perturbative one  is interpreted as
D-brane decay. It happens   that this model under some conditions
displays  a phantom behaviour \cite{AJK}. Note that
unlike phenomenological phantom models here phantom appears in an effective
theory. Since SFT  is a consistent theory this approach does not suffer from
usual problems
which are inevitable for phenomenological phantom models. The
UV completion is supposed to be solved by extending  the one mode (tachyon)
approximation.

SFT in the flat background   dictates a particular value of the D-brane tension.
It can be found from the requirement that the total energy of the system in the true non-perturbative vacuum is zero.
In cosmology this  total energy of the system in the true non-perturbative vacuum
can be interpreted as the cosmological constant. It has been conjectured that an existence
of a rolling solution describing a smooth transition to the true vacuum
does define the value of the cosmological constant \cite{IA}.
We cannot prove this conjecture but
arguments to it favor can be given using the local approximation \cite{AKV}.
A recent breakthrough in   solving numerically the full nonlinear and nonlocal  system
of equations \cite{LJ}
also supports this
expectation.

\section{Model}
Our model is given by the following  action \cite{IA}
$$
S=\int d^4x\sqrt{-g}\left(\frac{R}{2\kappa^2}+\frac1{\lambda_4^2}\left(-\frac{\xi^2\AP}{2}
g^{\mu\nu}\pd_{\mu}\phi(x)\pd_{\nu}\phi(x)+\frac1{2}\phi^2(x)-
\frac{1}{4}\Phi^4(x)-T^\prime\right)\right)
$$
Here
$g_{\mu\nu}$,  $\kappa$ and $\lambda_4$ are the four-dimensional metric, gravitational coupling constant
 and scalar field coupling constant, respectively;
$\frac1{\lambda_4^2}=\frac{v_6 M_s^4}{g_o}(\frac{M_s}{M_c})^6,$
$g_o$ is
the open string dimensionless coupling constant, $M_s$ is the string
scale $M_s=1/\sqrt{\alpha^\prime}$ and
 $M_c$ is a scale of the compactification,  $v_6$ is a number related with
 a volume of
 the  6-dimensional compact space. $T^\prime=1/4+\Lambda ^\prime,$ where
 $\Lambda ^\prime$ is a dimensionless
 cosmological constant. $\phi$ is a tachyon field and  $\Phi$ is related with
 $\phi$ by the following relation
 $
\Phi=e^{\frac{\AP}{8}\square_g}\phi,
~~~{\mbox {where}}~~~
\square_g=\frac1{\sqrt{-g}}\partial_{\mu}\sqrt{-g}g^{\mu\nu}\partial_{\nu}.$
This form of the nonlocal interaction is defined by the cubic fermionic SFT (CFSFT)
\cite{NPB}. More precisely,
the CFSFT  brings a more complicated form of the interaction, but by an analogy with
the flat case \cite{yar,AJK,LJ}
we believe that this  approximation catches essential physical properties of the model
   and
 $\xi^2\approx 0.9556$ is a constant defined by the CFSFT.

In the  spatially flat FRW metric
with   a  scale factor $a(t)$ the equation  for a space homogenous
tachyon field
$\Phi$
and the Friedmann equations  have the
form
~\cite{IA}
\begin{eqnarray}
\label{EOM_ST0approx_phi}
 \left(\xi^2{\cal D}+1\right)e^{-\frac{1}{4}{\cal D}}\Phi &=&\Phi ^3,\,\,\,\,\,\,\,\,\,\,\,
{\cal D}=-\pd _t^2-3H(t)\pd_t,\,\,\,\,\,H=\pd_t a/a\\\
\label{EOM_ST0approx}
3H^2&=&\frac{\kappa^2}{\lambda_4^2}\left(\frac{\xi^2}{2}\partial_t
\phi^2-\frac{1}{2}\phi^2+\frac{1}{4}\Phi^4+{\cal E}_1+{\cal E}_2+T^\prime\right),
\\
{\cal E}_{1}&=& - \frac{1}{8}\int_0^1 ds\left((\xi^2{\cal D}+1)\,\,\,
e^{\frac{s-2}{8}{\cal D}}
\Phi \,\right)\cdot
\left({\cal D}\,\,
e^{-\frac{1}{8} s {\cal D}} \Phi\right),\\
{\cal E}_2&=& -\frac{1}{8} \int_0^1 ds\left(
\partial_{t}(\xi^2{\cal D}+1)\,e^{\frac{s-2}{8}{\cal D}}
\Phi \right)\cdot
\left(\partial_t
e^{-\frac{1}{8}s{\cal D}} \Phi\right).
%\label{nonloc}
\end{eqnarray}
The non-local energy ${\cal E}_{1}$
plays the role of an extra potential term and ${\cal E}_{2}$ the role of the
kinetic term. Note that here we use a dimensionless time $t\to t \sqrt{\alpha^\prime}$.

\section{How we study our model and what we get}

Equations (\ref{EOM_ST0approx_phi}) and (\ref{EOM_ST0approx}) form a rather
complicated system of  nonlinear nonlocal equations for functions $\Phi$ and $H(t)$ because of the presence of an infinite
number of derivatives and a non-flat metric.
Before to discuss  the methods of study  this model let us mention the known
methods of study  equation (\ref{EOM_ST0approx_phi})
in the flat background, $H=0$,
\begin{equation}
\left(-\xi^2\partial_t ^2+1\right)e^{\frac{1}{4}\partial_t ^2}\Phi(t)=\Phi(t)^3.
\label{EOM_ST0approx_flat}
\end{equation}
Equation (\ref{EOM_ST0approx}) in the flat case describes the energy conservation \cite{AJK}.

A boundary
problem $\Phi(\pm \infty)=\pm 1$ for (\ref{EOM_ST0approx_flat}) has been studied using:
\begin{itemize}
\item
a numerical method \cite{yar}
based on an integral representation of (\ref{EOM_ST0approx_flat}); it is
 related with a diffusion equation method \cite{VS} which
 uses an auxiliary function of two variables $\Psi(r,t)$
that is the subject of a linear equation and $\Psi(\frac14,t)=\Phi(t)$;
\item
a decomposition on local fields \cite{AV,AK,AJV}; this method works well for linear equations
and has been used to study solutions to
(\ref{EOM_ST0approx_flat}) near vacuum $\pm 1$;
\item
existence theorems \cite{VV,Jouk,Prokh};
\item
  almost exact  solutions methods \cite{Forini,AJ}; the approach  \cite{Forini}
uses  a diffusion equation method.
\end{itemize}
 The following two  characteristic  properties of (\ref{EOM_ST0approx_flat}) have been obtained
 \begin{itemize}
\item
an existence of a  critical point $\xi^2_{\text{cr}}\approx 1.38$ such that for
$\xi^2<\xi^2_{\text{cr}}$
eq. (\ref{EOM_ST0approx_flat}) has  a rolling solution \cite{yar} interpolating between
$\pm 1$;
\item an existence of a dominance of an extra non-local
kinetic term  ${\cal E}_{2}$  over  the local kinetic one \cite{AJK}
and as a result, an appearance  of  a phantom behavior providing $w<-1$.
\end{itemize}
These result have been obtained using numerical calculations. It is very interesting to study the
problem analytically and  also try to find  approximate models admitting explicit solutions and having above mentioned properties.
They  could be  two or more components local models.\\

An investigation of    non-flat eqs.
  (\ref{EOM_ST0approx_phi}) and (\ref{EOM_ST0approx}) is essentially more complicated.
  The following methods are used:
\begin{itemize}
\item
A decomposition on local fields and a modification of the potential have been used in \cite{AKV,AJ,AKV2,AJV}.
A simplest one phantom mode approximation with an explicit form of the solution $\phi(t)=\tanh (t)$
is realized for a six-order potential \cite{AKV} and gives $H_0=1/3m^2_p$. Assuming that
$M_c\sim M_p$ and $M_s\sim 10^{-6.6}M_p$
we get
$$H_0\sim 10^{-60}M_p.$$
\item an analytic approach
that is closely related with
 the diffusion equation method \cite{Nardelli}.
\item
A numerical study has been performed in \cite{LJ}, where the diffusion equation method has been used  to define $\exp {\cal D}$
 and a
double-step iteration procedure has been proposed.
\end{itemize}
The following physical effects are found  in \cite{LJ}
%and (L.Joukovskaya, Pascos, July 07)
 \begin{itemize}
\item
For $\xi^2<\xi^2_{\text{cr}}\approx 1.18$ and  $\Lambda=\Lambda(\xi)$ the system
(\ref{EOM_ST0approx_phi}), (\ref{EOM_ST0approx}) has a rolling solution.
\item
For $\xi^2<\xi^2_{\text{shape}}$ and $t>0$ the Hubble function $H(t)$ is a
function which has small fluctuations about a monotonic function $H_l(t)$
with an asymptotic
 $H_0$; \\ for $\xi^2_{\text{shape}}<\xi^2<\xi^2_{\text{cr}}$ $H(t)$ describes fluctuations
 about a function $H_l(t)$ that has
   two maximum.
To
realize this approximated shape $H_l(t)$  by local fields one needs
at least two fields \cite{AKV2}.
\end{itemize}
As in the flat case it would be   very interesting  to find approximate analytical solutions which
exhibit these properties.
Note that two maximum shape regime for $H(t)$ is interesting in a context of building   an unified   cosmological
evolution. Let us also note that there are  applications of  non-local SFT models to inflation \cite{Lidsey,Cline}
and cosmological singularity (see refs. in \cite{AJV}).
 \begin{theacknowledgments}
  I thank the organizers of PASCOS for the invitation to a very stimulating conference.
  I am grateful to L.V.~Joukovskaya, A.S. Koshelev, S.Yu.~Vernov and
I.V.~ Volovich for collaboration and useful discussions.
The work  is supported in part by RFBR grant
05-01-00758, INTAS grant 03-51-6346 and Russian President's grant
NSh-2052.2003.1.
\end{theacknowledgments}

\bibliographystyle{aipproc}   % if natbib is available

\begin{thebibliography}{9}



\bibitem{IA} I.Ya. Aref'eva, AIP Conf. Proc. {\bf 826}
(2006) 301--311, astro-ph/0410443;
in: Contents and Structures of the Universe,
Proc. of the XLIst Rencontres de Moriond, 2006, pp. 131-135.

\bibitem{AMZ}
 I.Ya.~Aref'eva, P.B.~Medvedev, A.P.~Zubarev, Phys.Let. {\bf B} (1990);
 C.R.~Preitschopf, C.B.~Thorn, S.A.~Yost,  Nucl. Phys. {\bf B337} (1990) 363;
I.Ya.~Aref'eva, P.B.~Medvedev, A.P.~Zubarev, Nucl. Phys. {\bf B341} (1990)
464.
\bibitem{NPB} I.Ya.~Arefeva, D.M.~Belov, A.S.~Koshelev, P.B.~Medvedev,
 Nucl. Phys {\bf B638}(2002) 3.
 %% hep-th/0011117.
\bibitem{review-SFT} K.~Ohmori,   hep-th/0102085;
 I.Ya.~Aref'eva, D.M.~Belov, A.A.~Giryavets, A.S.~Koshelev,
P.B.~Medvedev, hep-th/0111208; W. Taylor,  hep-th/0301094.
\bibitem{AJK} I.Ya.~Aref'eva, L.V.~Joukovskaya, A.S.~Koshelev,
JHEP {\bf 0309} (2003) 012.
%% hep-th/0301137.

\bibitem{yar} Ya.I. Volovich,
 J. Phys. {\bf A36} (2003) 8685.
 %% math-ph/0301028.
\bibitem{LJ} L. Joukovskaya,\,
% Dynamics in Nonlocal Cosmological Models Derived from String Field Theory,
 arXiv:0707.1545;
 \,\, arXiv:0710.0404
% Rolling Tachyon in Nonlocal Cosmology
.
 \bibitem{AKV} I.Ya. Aref'eva, A.S. Koshelev, S.Yu. Vernov,
astro-ph/0412619; Phys. Lett. \textbf{B628} (2005) 1.



\bibitem{VS} V.S. Vladimirov, math-ph/0507018.

\bibitem{AV} I.Ya.~Aref'eva and  I.V. Volovich,  hep-th/0612098;
 Int. J. of Geom. Meth. Mod. Phys. V.4 (2007) 881-895,
hep-th/0701284.

\bibitem{AK} I.Ya.~Aref'eva, A.S.~Koshelev, JHEP 0702 (2007) 041;
 % hep-th/0605085;
%%CITATION = HEP-TH 0605085;%%
 A.S.~Koshelev, JHEP 0704 (2007) 029.
 % hep-th/0701103
\bibitem{AJV} I.Ya.~Aref'eva, L.V. Joukovskaya and S.Yu.Vernov,
JHEP 0707 (2007) 087.
% hep-th/0701184.
 \bibitem{VV} V.S. Vladimirov, Ya.I. Volovich, math-ph/0306018.
\bibitem{Jouk} L.V.~Joukovskaya,\,\,
%Theor. Math. Phys., {\bf 146} (2006) 335--342,
arXiv:0708.0642.
\bibitem{Prokh} D.V. Prokhorenko, math-ph/0611068.
%%CITATION = MAT-PH 0611068;%%

\bibitem{Forini}  V. Forini, G. Grignani, G. Nardelli, JHEP 0503 (2005) 079.
%hep-th/0502151.
\bibitem{AJ} I.Ya. Aref'eva and L.V. Joukovskaya,  JHEP 0510:087,2005.
%% hep-th/0504200
 \bibitem{Calcagni} G. Calcagni, JHEP \textbf{05} (2006) 012.
% hep-th/0512259
%%CITATION = HEP-TH 0512259;%%
\bibitem{AKV2} I.Ya. Aref'eva, A.S. Koshelev, S.Yu. Vernov, Phys.  Rev. \textbf{D72} (2005) 064017;
S.Yu. Vernov, astro-ph/0612487.
\bibitem{Nardelli}  G. Calcagni, M.Montobbio and G.Nardelli,
  arXiv: 0705.3043.



\bibitem{Lidsey} J.Lidsey, hep-th/0703007.
\bibitem{Cline} N. Barnaby, T. Biswas, J.M.  Cline,
hep-th/0612230; N. Barnaby,  J.M.  Cline, arXiv:0704.3426.




\end{thebibliography}

\end{document}